\documentclass[a4paper,fleqn,usenatbib]{mnras}
\usepackage[T1]{fontenc}
\usepackage{graphicx}

\title[PAHs and H$_2$ in the Ring Nebula]{Polycyclic aromatic hydrocarbons and molecular 
hydrogen in oxygen-rich planetary nebulae: the case of NGC\,6720}
\author[N.L.J. Cox et al.]{
N. L. J. Cox$^{1,2}$, 
P. Pilleri$^{1,2}$,
O. Bern\'e$^{1,2}$,
J. Cernicharo$^{3}$,
C. Joblin$^{1,2}$
\\ 
$^{1}$ Universit\'{e} de Toulouse, UPS-OMP, IRAP, 31028, Toulouse, France\\
$^{2}$ CNRS, IRAP, 9 Av. colonel Roche, BP 44346, F-31028 Toulouse, France\\
$^{3}$ Group of Molecular Astrophysics, ICMM, CSIC, C/Sor Juana In\'es de La Cruz N3, E-28049, Madrid, Spain}

\date{Accepted 2015 November 13. Received 2015 November 13; in original form 2015 August 31}
\pubyear{2015}
\begin{document}
\label{firstpage}
\pagerange{\pageref{firstpage}--\pageref{lastpage}}
\maketitle

\begin{abstract}
Evolved stars are primary sources for the formation of polycyclic aromatic hydrocarbons (PAHs) and dust grains. 
Their circumstellar chemistry is usually designated as either oxygen-rich or carbon-rich, although dual-dust 
chemistry objects, whose infrared spectra reveal both silicate- and carbon-dust features, are also known. 
The exact origin and nature of this dual-dust chemistry is not yet understood. \emph{Spitzer}-IRS  mid-infrared 
spectroscopic imaging of the nearby, oxygen-rich planetary nebula NGC\,6720 reveals the presence of the 
11.3~$\mu$m aromatic (PAH) emission band. It is attributed to emission from neutral PAHs, since no band
is observed in the 7--8~$\mu$m range. The spatial distribution of PAHs is found to closely follow that of the 
warm clumpy molecular hydrogen emission. Emission from both neutral PAHs and warm H$_2$ is likely to arise 
from photo-dissociation regions associated with dense knots that are located within the main ring. 
The presence of PAHs together with the previously derived high abundance of free carbon (relative to CO) 
suggest that the local conditions in an oxygen-rich environment can also become 
conducive to in-situ formation of large carbonaceous molecules, such as PAHs, via a bottom-up chemical pathway. 
In this scenario, the same stellar source can enrich the interstellar medium with both oxygen-rich dust 
and large carbonaceous molecules.
\end{abstract}


\section{Introduction}

The asymptotic giant branch (AGB) and planetary nebula (PN) phases of stellar evolution govern the chemical 
enrichment of the interstellar medium (ISM) by low- to intermediate-mass stars. Evolved stars are efficient dust 
factories, but the processes involved in dust formation and evolution are still shrouded in mystery. Unlike molecules, 
dust grains can only be efficiently formed in the innermost warm regions of evolved stars (\citealt{1988A&A...206..153G}). 
Carbon-rich evolved stars are formation sites of polycyclic aromatic hydrocarbons (PAHs).
Bottom-up formation is either via pyrolysis (\citealt{1989ApJ...341..372F}, \citealt{1992ApJ...401..269C}) or 
photolysis (\citealt{2004ApJ...608L..41C}).  The top-down process could occur through shock-, photo-,
or chemical destruction of carbonaceous or SiC grains (\citealt{1996ApJ...469..740J}, \citealt{2012A&A...542A..69P}, 
\citealt{2014NatCo...5E3054M}). 

Generally, silicate dust formation is related to oxygen/O-rich winds (C/O$<$1), while carbonaceous dust formation occurs 
primarily in carbon/C-rich winds (C/O$>$1). PAH features are present in PNe with C/O abundance ratios as low as $\sim$0.6 
(\citealt{2005MNRAS.362.1199C,2014ApJ...784..173D}). However, in several cases both oxygen-rich and carbon-rich dust 
features are seen in (spatially integrated) infrared spectra of PNe (e.g. \citealt{2009A&A...495L...5P}, \citealt{2009ApJ...703..585C}). 
\citet{2013A&A...558A.122G} found that dual-dust chemistry PNe with O-rich nebulae show weak PAH 
bands and crystalline/amorphous silicates, while the C-rich ones have strong PAH bands and very weak 
crystalline silicate features. \citet{2014A&A...567A..12G} identified the presence of dual-dust chemistry in PNe 
mainly with high-metallicity and relatively high-mass ($\sim$3--5~M$_\odot$) young PNe. Dual-dust chemistry 
in C-rich PNe has been explained in terms of different evolutionary phases.
The silicate dust emission is formed earlier. 
PAHs are formed in a later mass-loss phase and are present in the outflows (\citealt{2004ApJ...604..791M}). 

For PAHs observed in O-rich Galactic Bulge PNe (\citealt{2009A&A...495L...5P,2011MNRAS.414.1667G,2014MNRAS.441..364G}) 
this scenario seems less satisfactory since these PNe are mostly old, 
low-mass stars that are not expected to go through a third dredge-up phase, and therefore show no enhanced C/O ratios. 
An alternative explanation is that PAHs in these objects are formed in-situ in an UV-irradiated dense torus 
as suggested by \citet{2004ApJ...604..791M} and \citet{2004ApJ...608L..41C}. 
In this scenario PAHs form in a region just outside where some UV photons penetrate and dissociate 
CO and the free carbon can subsequently aggregate there and lead to formation of PAHs (e.g. \citealt{2004ApJ...604..791M}). 
\citet{2004ApJ...608L..41C} has shown that in dense PDRs of proto-PNe the photodissociation of CO and other species leads to the 
growth of carbon chains, and to the production of carbon clusters. This rich photochemistry suggests a possible bottom-up scenario 
for the formation of PAHs. \citet{2008A&A...483..831A} have shown that the same applies to O-rich environments as those found in 
protoplanetary disks where C$_2$H$_2$ and HCN are found with very high abundances.

In this Letter we present \emph{Spitzer} spectroscopy of the nearby PN NGC\,6720 (Ring Nebula). Sect.~\ref{sec:2} presents 
the \emph{Spitzer} observations of NGC\,6720. The results we obtained for the PAH and H$_2$ emission are discussed in 
Sect.~\ref{sec:3} and put in perspective in Sect.~\ref{sec:4}.

\begin{figure}
\centering
\includegraphics[width=0.85\columnwidth]{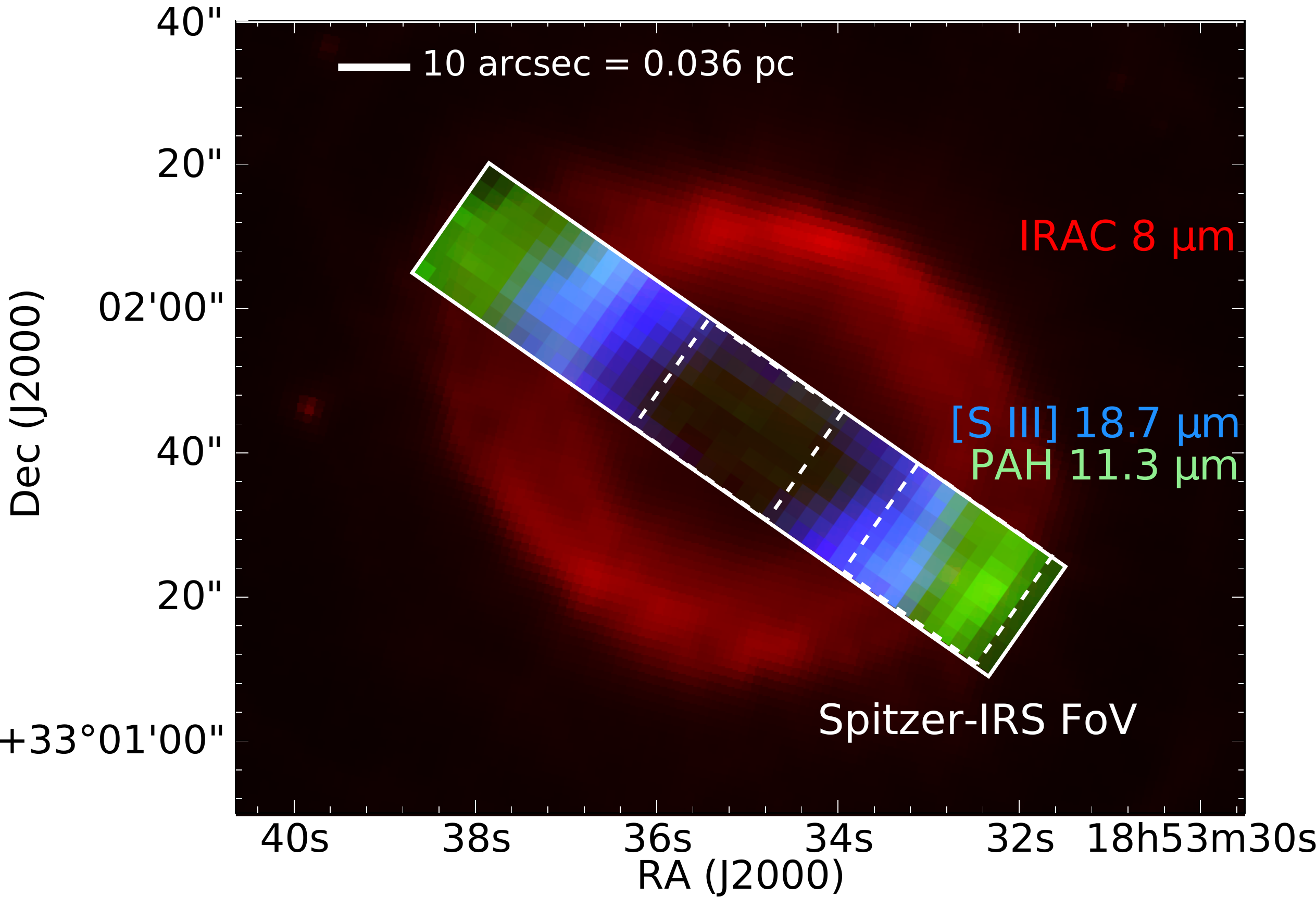}
\caption{
\emph{Spitzer}-IRS maps of the 11.3~$\mu$m emission feature (green) and the [S\,{\sc iii}] emission line (blue) in the 
Ring Nebula overlaid on top of the \emph{Spitzer} IRAC 8~$\mu$m image (red). The two white squares (dashed lines) 
indicate the extraction region for the spectra shown in Fig.~\ref{fig:1b}.}
\label{fig:1a}
\end{figure}

\begin{figure}
\centering
\includegraphics[width=1.05\columnwidth]{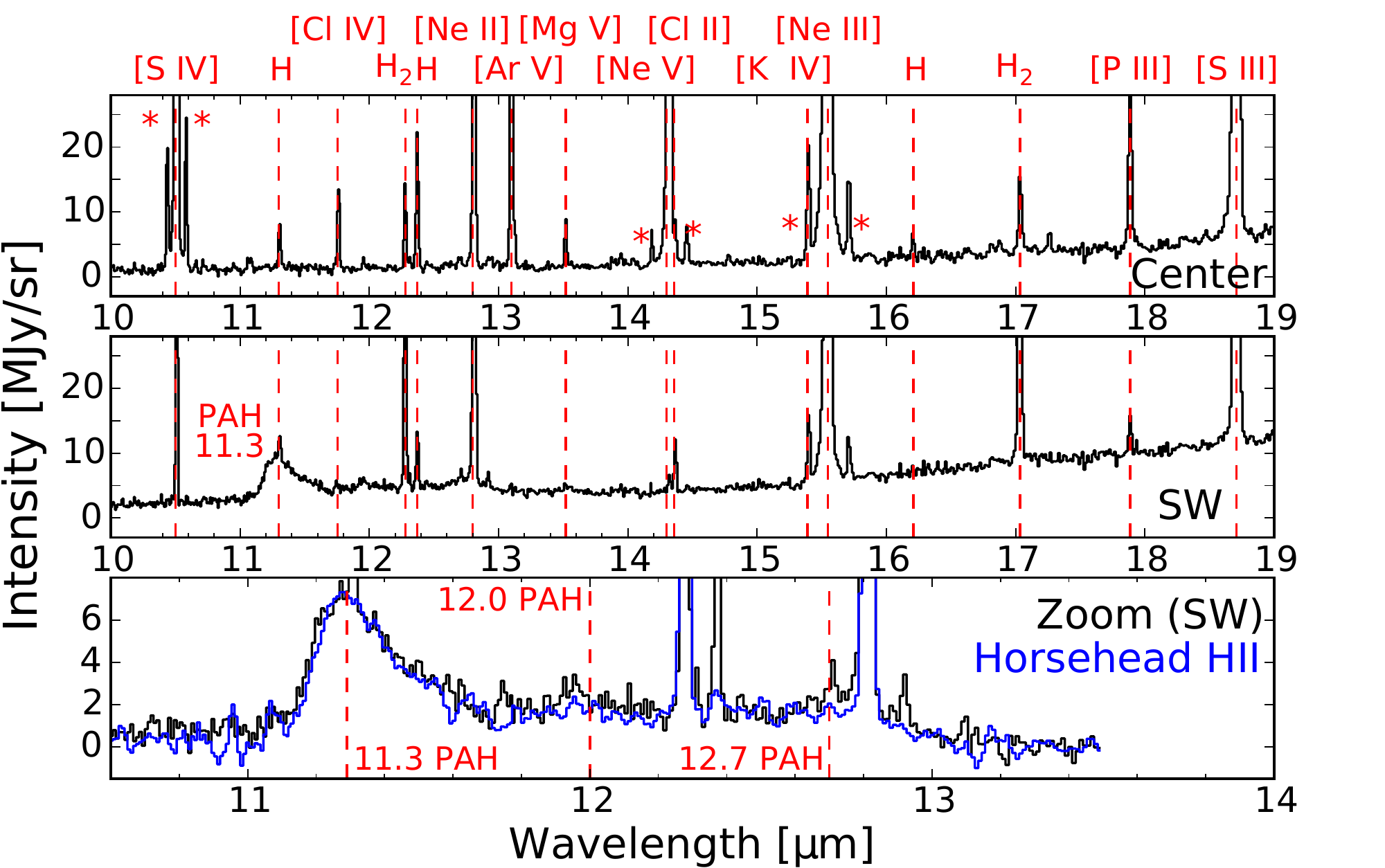}
\caption{
\emph{Spitzer}-IRS SH spectra extracted from the 3D spectral-cube at two positions in the Ring Nebula. 
The positions are indicated by the two white-dashed squares in Fig.~\ref{fig:1a}.
The top panel shows a spectrum from the central hot-gas cavity and the middle panel shows the spectrum ``SW'' 
arising from the main dust ring.  
Emission lines arising from several atoms and molecular hydrogen are labelled (ghost lines are indicated with asterisks).
The bottom panel shows a close-up view of the  (continuum subtracted) 11--13~$\mu$m PAH emission complex. 
The PAH emission spectrum of the Horsehead H\,{\sc ii} region (\citealt{2007A&A...471..205C}), 
scaled by a factor of 0.4 to match the 11.3~$\mu$m peak intensity, is shown for reference.}
\label{fig:1b}\label{fig:5}
\end{figure}

\section[]{Spectroscopic imaging of NGC\,6720}\label{sec:2}

NGC\,6720 is a nearby (740$^{+400}_{-200}$~pc; \citealt{2009AJ....137.3815O}),  oxygen-rich bipolar PN. 
Its main ring corresponds to an ionisation bounded torus structure 
seen nearly pole-on with lobes directed along the line-of-sight (c.f. \citealt{2013AJ....145...92O}). 
The central star of the PN (CSPN) has an effective temperature of $T_\mathrm{eff}$ = 120\,000~K and a 
stellar luminosity $L_\star = 200$~$L_\odot$ (\citealt{2007AJ....134.1679O}). The reported C/O abundance 
ratio ranges from 0.4 to 1.0 (\citealt{2014ApJ...784..173D}). 

NGC\,6720 has been observed with the \emph{Spitzer} Space Telescope (\citealt{2004ApJS..154....1W}) as 
part of program 40536 (PI: H.\,Dinerstein). The IRS-SH observations (\citealt{2004ApJS..154...18H}) have 
been processed with CUBISM (\citealt{2007PASP..119.1133S}), including sky background subtraction and 3-sigma clipping.
The 3D spectral data-cube allows to study the mid-infrared spectrum (10--19.6~$\mu$m, R$\sim$600) of NGC\,6720 
in a region of 99~$\times$~18 arcsec with a plate scale of 1.8 arcsec per pixel, as shown in Fig.~\ref{fig:1a}.

In an initial assessment we extracted two (average) spectra, one towards the central star and one pertaining to 
the main dust ring. The extracted spectra are shown in Fig.~\ref{fig:1b}. Both spectra clearly differ in terms of 
relative strength of the gas emission lines.  All the gas emission lines typically detected between 10--20~$\mu$m 
for PNe are present in NGC\,6720. Most noticeable, however, is the presence of the 11.3~$\mu$m PAH band. 
The 12.0 and 12.7~$\mu$m PAH bands are also detected in the spectrum for the main ring, confirming the presence of PAHs. 
The presence of the 11.3~$\mu$m PAH band is further corroborated by lower-sensitivity IRS SH spectra (calibration program 1424) 
taken at two positions (N-S) on the ring perpendicular to the data cube (E-W). Previously, PAH emission bands were thought to 
be absent in this PN, although in hindsight the 11.3~$\mu$m band (contaminated 
by a H\,{\sc i} line) is tentatively seen in the \emph{Spitzer}-SL spectra shown by \citet{2009eimw.confE..29H}. 

No clear detection of a PAH feature in the 7--8~$\mu$m range could be obtained; The data is very noisy and suffers 
from a bad overlap between the two SL modules of the IRS spectrograph.
Previous spectra obtained with ISOCAM CVF have insufficient spectral resolution and sensitivity to detect PAH bands 
in the presence of strong atomic emission lines. The ISO-SWS aperture does not cover the main ring.
\emph{Spitzer} LH/LL (20--40~$\mu$m) spectroscopy (program 1424) of the dust ring is of insufficient sensitivity and spectral
resolution to discern the presence of any dust features.

\begin{figure}
\centering
\includegraphics[bb=0 15 330 500,clip,width=0.83\columnwidth]{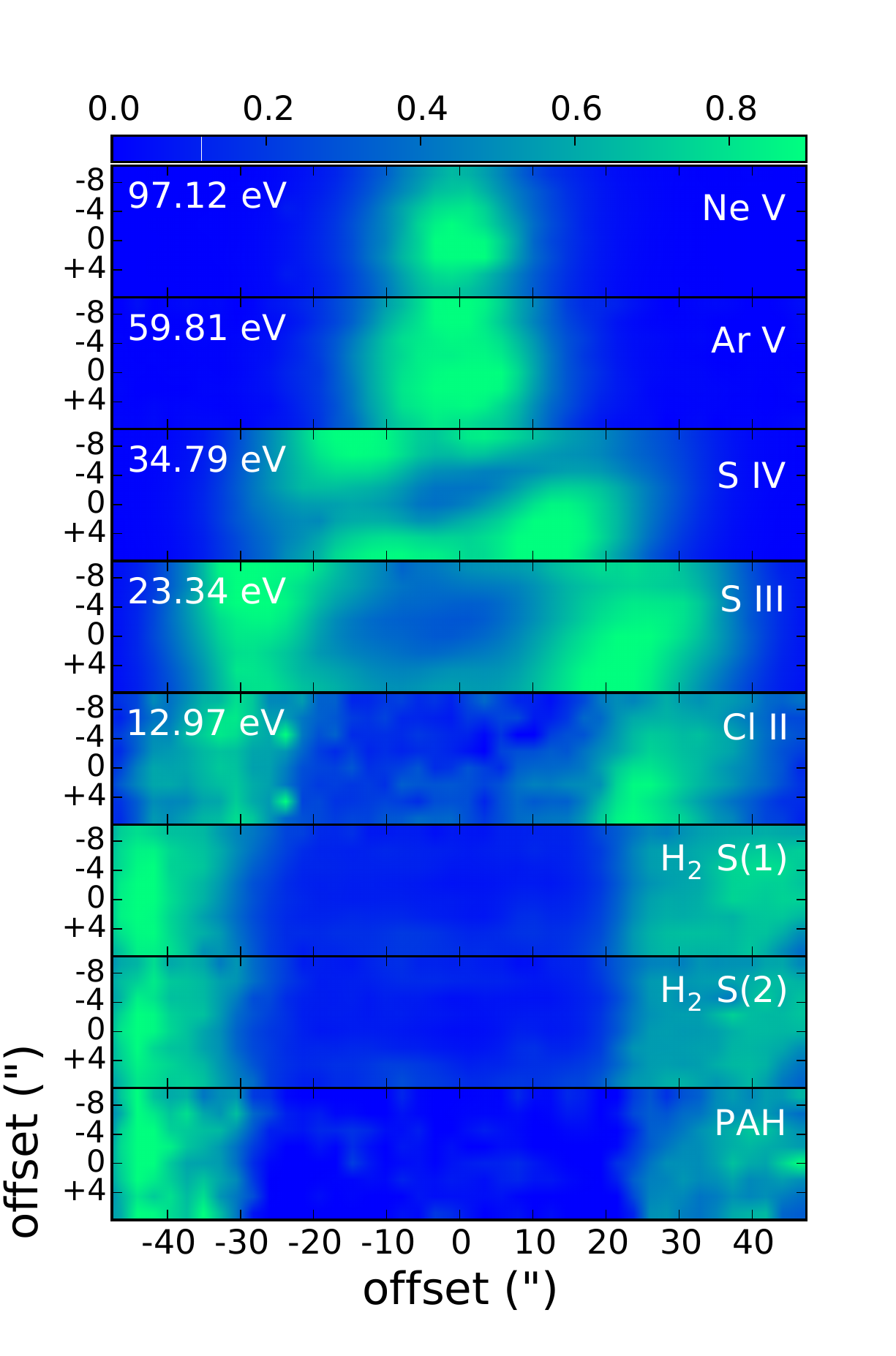}
\caption{2D spatial distribution of line intensity for the atomic (Ar\,{\sc v}, S\,{\sc iv}, S\,{\sc iii}, Cl\,{\sc ii}, H\,{\sc i}) 
and molecular (H$_2$, PAH) tracers in NGC\,6720. Intensities are normalised to the maximum value found for each species.
Negative offsets are towards the south-west. 
The atomic species and the energies (eV) required to bring them in the specified ionisation state are indicated.}
\label{fig:2}
\end{figure}

Line intensities for a representative subset of species are extracted from the \emph{Spitzer} spectrum for each position 
in the spectral map. This set of lines traces roughly the ionising radiation field throughout the nebula, that corresponds to 
the photon energies ($\sim$13--60~eV) required to bring atoms in a certain ionisation state. The spatial distribution of 
the mid-infrared H$_2$ S(1) and S(2) emission lines at 17.04 and 12.28~$\mu$m, and the 11.3~$\mu$m PAH band is 
shown in Fig.~\ref{fig:2}. 

\section{Discussion}\label{sec:3}

\begin{figure}
\includegraphics[width=1.\columnwidth]{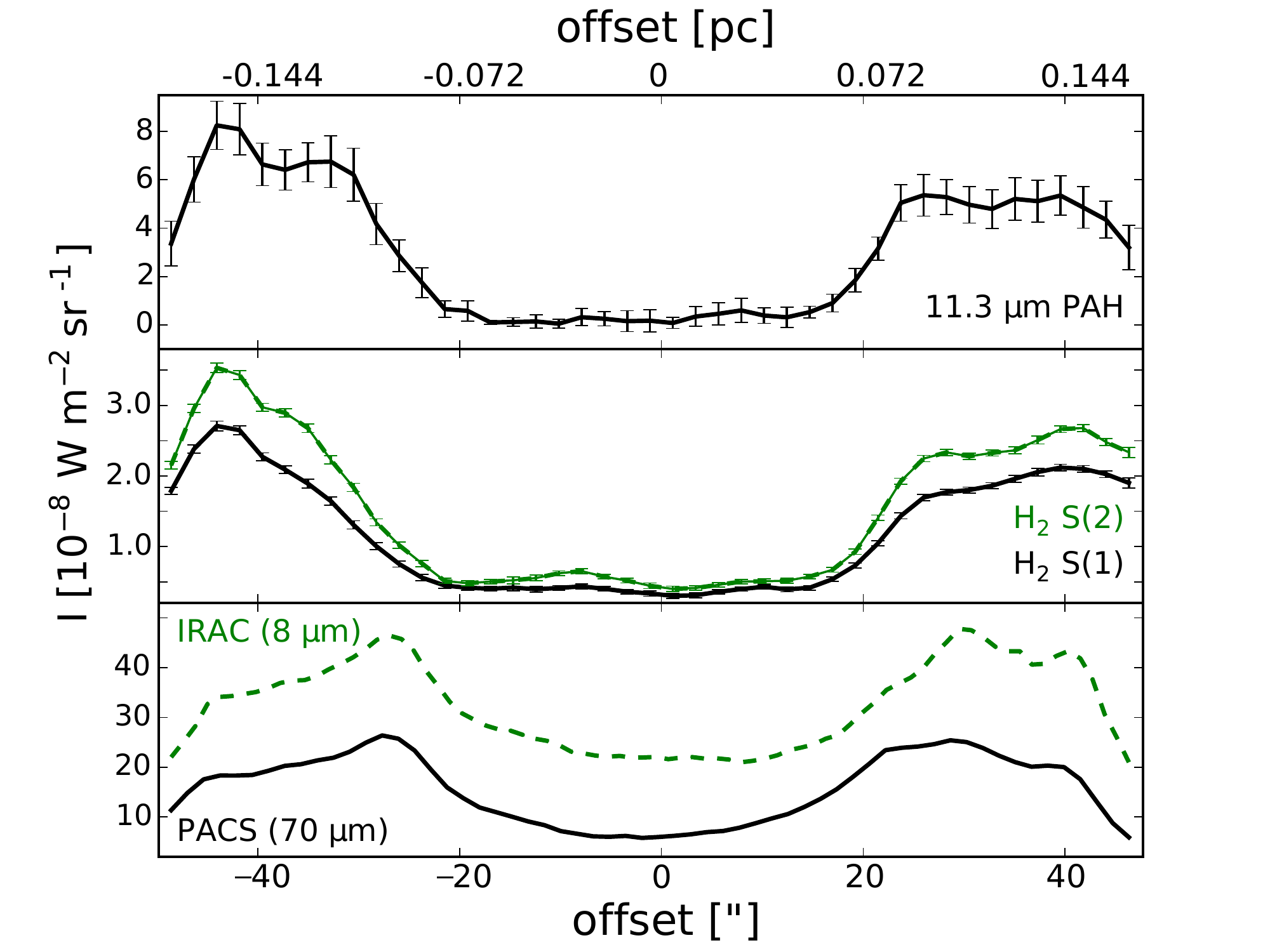}
\caption{Radial profiles for PAH and H$_2$ integrated intensities, and infrared (dust) emission at 8 and 70~$\mu$m (W\,m$^{-2}$\,sr$^{-1}$).}
\label{fig:2b}\label{fig:3}
\end{figure}

Fig.~\ref{fig:3} reveals a close relation between the radial distributions of PAH, H$_2$, and dust emission in NGC\,6720 
(see \citealt{2010A&A...518L.137V} for a comparative study between H$_2$ and dust emission). 
The co-spatial distribution of H$_2$ and PAH emission shows that both species are present and excited at similar 
locations within the molecular ring. PAHs could also be involved in the formation of H$_2$
(\citealt{2003A&A...397..623H}, \citealt{2015A&A...579A..72B}). 
Next, we discuss the observed PAH bands (Sect.~\ref{sec:pah}) and give a brief analysis of the H$_2$ 
emission (Sect.~\ref{sec:H2}). Sect.~\ref{sec:hydrocarbon} then examines these results in the context of hydrocarbon photochemistry.

\subsection{PAH emission bands: profile and intensity}\label{sec:pah}

The 11.3~$\mu$m PAH band is observed at the position of 11.26 $\pm$ 0.05~$\mu$m, which is typical for that band 
(\citealt{2001A&A...370.1030H}). The full-width-at-half-maximum is $\sim$0.33~$\mu$m.
This is larger than typical, and is due to a red wing extending up to $\sim$11.6~$\mu$m. 
The total integrated emission of the 11.3~$\mu$m feature in the main ring ranges from $\sim$2 to 8\,$\times$\,10$^{-8}$~W\,m$^{-2}$\,sr$^{-1}$. 
For the PAH bands extracted from the ``SW'' region (Figs.~\ref{fig:1b}) the peak intensity is $\sim$7~MJy\,sr$^{-1}$ 
with an integrated line intensity of 6.7\,$\times$\,10$^{-8}$~W\,m$^{-2}$\,sr$^{-1}$.
I(11.3~$\mu$m), normalised by the total integrated infrared emission (TIR), is consistent  
with the relation with C/O ratio as derived by \citet{2005MNRAS.362.1199C} for a large sample of PNe with C/O$\approx$0.3--3.
The integrated line intensities of the 12.0 and 12.7~$\mu$m PAH bands in the ``SW'' region are estimated 
to be $2.1 \pm 0.3$ and $2.2 \pm 0.7$\,$\times$\,10$^{-8}$~W\,m$^{-2}$\,sr$^{-1}$, respectively.
The 7.7~$\mu$m PAH feature is not detected (Sect.~\ref{sec:2}). 
We derive an upper limit of its peak intensity, $I(7.7)=0.3 \pm 1.3$\,MJy\,sr$^{-1}$, to be compared to $I(11.3)=7.0 \pm 0.82$\,MJy\,sr$^{-1}$.

\citet{2002A&A...388..639P} modelled the emission of a distribution of PAHs in IRAS\,21282+5050 suggesting that 
the extended profile of the 11.3~$\mu$m band is due to the smallest sizes of the PAH distribution (containing between 30 and 48 carbon atoms).
\citet{2015MNRAS.448.2960C} also modelled the 11.3~$\mu$m emission band with neutral PAHs. These authors conclude 
that a profile with a peak at a relatively longer wavelength and with a more extended red tail (as well as a less steep short 
wavelength side) implies that lower-mass PAHs are relatively more abundant. 
\citet{2001A&A...370.1030H} argue that the I(12.7)/I(11.3) band strength ratio is determined by molecular structure, with low 
values associated to (compact) PAHs present in PNe.
The I(7.7)/I(11.3) ratio is $<$ 0.26 which is below the values determined by \citet{2012A&A...542A..69P} for PAH and 
PAH$^+$ template spectra, which are 0.56 and 2.77, respectively. Although such a low value is rather exceptional, 
quantum chemistry calculations indicate that it is possible for neutral PAHs (\citealt{Langhoff1996}; see 
\citealt{2011EAS....46...75P} for a review) and even for very large ones (\citealt{2008ApJ...678..316B}; \citealt{2012ApJ...754...75R}). 

Ionised PAHs are expected to dominate in PDRs with large values of $G_0/n_e > 10^4$ (where $G_0$ is the radiation field 
strength in Habing units of 1.6\,10$^{-6}$ W\,m$^{-2}$), whereas mostly neutral PAHs or even negatively charged PAHs are expected 
to be found in PNe and H\,{\sc ii} regions when the medium is fully ionised ($n_e \sim n_H$) and low values of 
$G_0/n_e < 10^3$ prevail (\citealt{2008A&A...490..189J}). For the ring of NGC\,6720 the effective G$_0$ is 
$\approx$200--400\footnote{From the combined \emph{Herschel}-PACS (\citealt{2010A&A...518L.137V}) and 
\emph{Spitzer}-LH dust emission, the total integrated infrared dust emission, TIR~$\approx$~2.0~$\times$~10$^{-5}$~W\,m$^{-2}$\,sr$^{-1}$, is a 
tracer of the energy input: G$_0$(TIR) = $4 \pi\ \times$~TIR / ($1.6 \times 10^{-6}$) $\approx 180$. 
Alternatively the radiation field strength at 0.1~pc from the central star can be estimated as from an on-the-spot approximation, 
G$_0$ = 625 L$_\star \chi / (4 \pi d^2) = 402$ (\citealt{2005pcim.book.....T}, p.319).}.
A predominance of neutral PAHs has previously been reported for the Horsehead nebula H\,{\sc ii} region, where $G_0 \sim 100$ 
(\citealt{2007A&A...471..205C}). 
In this case the usually strong 6.2 and 7.7~$\mu$m PAH bands are also absent, whereas the 11--13~$\mu$m PAH emission 
spectrum is very similar to that of NGC\,6720 (Fig.~\ref{fig:5}, bottom panel; \citealt{2007A&A...471..205C}).
Similar profiles are also seen in the PNe IRAS\,21282+5050 and IRAS\,17047-5650 (\citealt{2001A&A...370.1030H}) 
and mixed-chemistry post-AGB star IRAS\,16279-4756  (\citealt{2004ApJ...604..791M}).

\subsection{H$_2$ temperature and density}\label{sec:H2}

The presence of H$_2$ in PNe is attributed to PDRs associated with the dense knots residing 
in the ionized regions (\citealt{2003PASP..115..170S, 2015ApJ...808..115M}).
\citet{2011A&A...528A..74A} find that H$_2$ infrared emission lines in PNe are produced mainly in the warm 
and partially ionized transition zone between the ionised H\,{\sc ii} region and the neutral PDR. 
Additional heating and ionisation of the gas is required to explain the strong H$_2$ emission lines.
Possible mechanisms are soft X-rays from the hot CSPN (\citealt{1998A&A...337..517N}, 
\citealt{1999A&A...342..823V}, \citealt{2011A&A...528A..74A}; though no diffuse X-ray emission is observed; \citealt{2012AJ....144...58K}), shocks (\citealt{2001AJ....122.3293A}),
or the development of an extreme-UV dominated advective PDR in which the ionisation and dissociation fronts 
have merged (\citealt{2007ApJ...671L.137H}).

The integrated line strengths for the H$_2$(0-0) S(1) and S(2) emission from the main ring 
range from 1 to~3\,$\times 10^{-8}$~W\,m$^{-2}$\,sr$^{-1}$ (Fig.~\ref{fig:2b}). In addition, we extracted H$_2$ S(1) to S(4) 
integrated line intensities from a small region in the north-east part of the ring for which a single (lower resolution) 
\emph{Spitzer}-SL spectrum is available.  Assuming thermal equilibrium we use the line intensities to derive the
rotational temperature and column density of H$_2$: T(H$_2$) = $620^{+121}_{-87}$~K, N(H$_2$) = $1.0^{+0.6}_{-0.4}$~$\times$~10$^{18}$~cm$^{-2}$. 
These values are comparable to those derived by \citet{1998ApJ...495L..23C} for the Helix Nebula: T(H$_2$) = 900~K and N(H$_2$) = 2.0 $\times$ 10$^{18}$~cm$^{-2}$.

The H\,{\sc ii} region drives a shock in the globule and hence the pressure in the external molecular shell around the globule 
(where H$_2$ is emitting) will be comparable to the pressure in the H\,{\sc ii} region (e.g. \citealt{1989ApJ...346..735B}). 
Pressure equilibrium at the ionisation front is given by $2\,n_e\,T_e = n_H\,T_\mathrm{gas}$. If photo-evaporation occurs this 
provides a lower density limit. 
Assuming that the H$_2$ rotational temperature we derived above is an upper limit to the gas kinetic 
temperature (i.e. v=0 rotational levels are excited by collisions but also by UV fluorescence cascades in the surface PDR of the globules),
as well as $T_e = 10\,500 \pm 1000$~K (\citealt{2004MNRAS.353.1251L}) and 
$n_e = 400 \pm 200$~cm$^{-3}$ (\citealt{2013AJ....145...93O}), we obtain $n_H > 1$--2.5\,$\times$\,10$^4$~cm$^{-3}$.
This is consistent with estimates from optical studies  (\citealt{2002AJ....123.3329O}).

The gradual increase of H$_2$ intensity from 0.07 to 0.14~pc is linked to the decrease of the H$_2$ 
destruction rate with the attenuation of UV photons (corresponding to $A_V \approx 0.5$; \citealt{2012A&A...542A..69P}), 
while the sharp increase of PAH intensity at $\sim$0.08~pc is primarily due to a density gradient 
(PAHs are less sensitive to destruction by UV photons than H$_2$). 
$A_V = 0.5$~mag corresponds to N(H)$\sim$ 10$^{21}$~cm$^{-2}$, which yields an \emph{average} volume density of 10$^{4}$~cm$^{-3}$, with
higher densities inside the knots.

\subsection{Hydrocarbon photochemistry}\label{sec:hydrocarbon}

We expand here on the scenario proposed by e.g. \citet{2004ApJ...604..791M}, \citet{2004ApJ...608L..41C}, and 
\citet{2011MNRAS.414.1667G} in which CO is dissociated and PAHs are formed from free carbon. In the case of 
NGC\,6720 the abundance of elemental carbon (i.e. C and C$^+$) is an order of magnitude higher than that of CO
(\citealt{1994A&A...281L..93B}, \citealt{2012A&A...542L..20S}), likely a result of dissociation of the latter by the 
strong radiation field of the high-excitation CSPN. Both CO and C emitting regions are clumped. As shown in 
the previous section these clumps have high density ($>10^4$~cm$^{-3}$) and are exposed to a moderate radiation 
field ($G_0 \sim 200$). \citet{2004ApJ...608L..41C} and \citet{2008A&A...483..831A} have included in their chemical 
models for C and O-rich environments reactions such as 
C$_2$ + H$_2$ $\rightarrow$ C$_2$H + H, C$_2$H + H$_2$ $\rightarrow$ C$_2$H$_2$ + H, and C$^+$ + H$_2$ = CH$^+$ + H 
which have moderate activation barriers, which make them very slow in the cold interstellar medium, but become 
rapid enough to control the abundance of C-bearing species at the high temperatures and high H$_2$ densities prevailing in the PDRs of (proto-)PNe. 
This is because chemical reactions involving H$_2$ v=0 with radicals in a warm gas become significantly
efficient if the volume density is large enough (\citealt{2004ApJ...608L..41C}). 
H$_2$ v=1 reactions with more complex organic molecules could also overcome activation barriers of several thousands Kelvin
allowing the growth of chemical complexity.
In a C-rich PDR environment C$_2$H$_2$ and HCN play a key role in the photochemistry (\citealt{2004ApJ...608L..41C}), 
while for O-rich environments \citet{2008A&A...483..831A} have shown that the photodissociation of CO and N$_2$ provide 
the path to the formation of C$_2$H$_2$ and HCN with abundances similar to those of C-rich environments. 
The high (H$_2$) gas temperature, together with freely available C, suggests a very fast and rich photochemistry at the 
surface of the dense knots existing in the main ring of NGC\,6720. Together, these conditions favour a bottom-up 
formation of large carbonaceous molecules.

It is not clear how this process would give rise to a peculiar PAH population such as observed for NGC\,6720. 
The PAH size distribution will be the result of the equilibrium between formation and destruction, where UV destruction 
generally favours increased abundance of large PAHs which are more stable (\citealt{2013A&A...552A..15M}).  
Although the large PAHs have high probability to be ionized rather than photo-dissociated (\citealt{2015ApJ...804L...7Z}), 
the high electron abundance provides a competing recombination rate, which shifts the balance in favour of neutral PAHs. 

\section{Summary \& Conclusions}\label{sec:4}

We present the unexpected detection of a weak 11.3~$\mu$m PAH emission
band in the main dust torus of O-rich PN NGC\,6720. 
The band profile (position and red wing) and the absence of noticeable bands in the 7--8~$\mu$m range
indicates that the emitting population consists primarily of neutral PAHs and contains a 
relatively higher abundance of smaller specie as found in most other objects.
The spatial distribution of PAHs is closely correlated with the clumpy distribution of H$_2$
revealing the presence of PDRs associated with dense knots
exposed to a moderate radiation field.
For the physical environment associated with dense knots inside an ionised region,
chemical models provide a way to transform a gas dominated by
H$_2$, CO, H$_2$O, N$_2$ with C/O$<$1 into a gas where complex carbon chemistry
can occur and this might provide a pathway to form PAHs through a bottom-up process.
In NGC\,6720 reactions with large activation barriers could be facilitated by excitation of H$_2$ in v=1.
A more detailed inventory of the properties and chemistry of nebular knots in oxygen-rich PNe 
will require observations at higher angular resolution.
Regardless of the exact formation mechanism involved, our analysis indicates that complex hydrocarbon chemistry, 
including the formation of large molecules (with more than 30 C atoms), can occur in PDRs associated with
high-density knots located within the ionized regions of O-rich PNe. 
Therefore, if knots are commonly formed in (old) O-rich PNe, these can enrich the ISM also with carbonaceous material.

\section*{Acknowledgments}

We acknowledge helpful comments by the referee.
The research leading to these results has received funding from the European Research Council under the European 
Union's Seventh Framework Programme (FP/2007-2013)  ERC-2013-SyG, Grant Agreement n. 610256 NANOCOSMOS. 
PP acknowledges financial support from the Centre National Etudes Spatiales (CNES). 
The IRS was a collaborative venture between Cornell University and Ball Aerospace Corporation funded by NASA 
through the Jet Propulsion Laboratory and Ames Research Centre.

\footnotesize{
 \bibliographystyle{mnras}
 \bibliography{cox_ngc6720_bibtex}

\begin{thebibliography}{}
\makeatletter
\relax
\def\mn@urlcharsother{\let\do\@makeother \do\$\do\&\do\#\do\^\do\_\do\%\do\~}
\def\mn@doi{\begingroup\mn@urlcharsother \@ifnextchar [ {\mn@doi@}
  {\mn@doi@[]}}
\def\mn@doi@[#1]#2{\def\@tempa{#1}\ifx\@tempa\@empty \href
  {http://dx.doi.org/#2} {doi:#2}\else \href {http://dx.doi.org/#2} {#1}\fi
  \endgroup}
\def\mn@eprint#1#2{\mn@eprint@#1:#2::\@nil}
\def\mn@eprint@arXiv#1{\href {http://arxiv.org/abs/#1} {{\tt arXiv:#1}}}
\def\mn@eprint@dblp#1{\href {http://dblp.uni-trier.de/rec/bibtex/#1.xml}
  {dblp:#1}}
\def\mn@eprint@#1:#2:#3:#4\@nil{\def\@tempa {#1}\def\@tempb {#2}\def\@tempc
  {#3}\ifx \@tempc \@empty \let \@tempc \@tempb \let \@tempb \@tempa \fi \ifx
  \@tempb \@empty \def\@tempb {arXiv}\fi \@ifundefined
  {mn@eprint@\@tempb}{\@tempb:\@tempc}{\expandafter \expandafter \csname
  mn@eprint@\@tempb\endcsname \expandafter{\@tempc}}}

\bibitem[\protect\citeauthoryear{{Ag{\'u}ndez}, {Cernicharo}  \&
  {Goicoechea}}{{Ag{\'u}ndez} et~al.}{2008}]{2008A&A...483..831A}
{Ag{\'u}ndez} M.,  {Cernicharo} J.,   {Goicoechea} J.~R.,  2008, \mn@doi [\aap]
  {10.1051/0004-6361:20077927}, \href
  {http://adsabs.harvard.edu/abs/2008A%26A...483..831A} {483, 831}

\bibitem[\protect\citeauthoryear{{Aleman} \& {Gruenwald}}{{Aleman} \&
  {Gruenwald}}{2011}]{2011A&A...528A..74A}
{Aleman} I.,  {Gruenwald} R.,  2011, \mn@doi [\aap]
  {10.1051/0004-6361/201014978}, \href
  {http://adsabs.harvard.edu/abs/2011A%26A...528A..74A} {528, A74}

\bibitem[\protect\citeauthoryear{{Arias}, {Rosado}, {Salas}  \&
  {Cruz-Gonz{\'a}lez}}{{Arias} et~al.}{2001}]{2001AJ....122.3293A}
{Arias} L.,  {Rosado} M.,  {Salas} L.,   {Cruz-Gonz{\'a}lez} I.,  2001, \mn@doi
  [\aj] {10.1086/324446}, \href
  {http://adsabs.harvard.edu/abs/2001AJ....122.3293A} {122, 3293}

\bibitem[\protect\citeauthoryear{{Bachiller}, {Huggins}, {Cox}  \&
  {Forveille}}{{Bachiller} et~al.}{1994}]{1994A&A...281L..93B}
{Bachiller} R.,  {Huggins} P.~J.,  {Cox} P.,   {Forveille} T.,  1994, \aap,
  \href {http://adsabs.harvard.edu/abs/1994A%26A...281L..93B} {281, L93}

\bibitem[\protect\citeauthoryear{{Bauschlicher}, {Peeters}  \&
  {Allamandola}}{{Bauschlicher} et~al.}{2008}]{2008ApJ...678..316B}
{Bauschlicher} Jr. C.~W.,  {Peeters} E.,   {Allamandola} L.~J.,  2008, \mn@doi
  [\apj] {10.1086/533424}, \href
  {http://adsabs.harvard.edu/abs/2008ApJ...678..316B} {678, 316}

\bibitem[\protect\citeauthoryear{{Bertoldi}}{{Bertoldi}}{1989}]{1989ApJ...346..735B}
{Bertoldi} F.,  1989, \mn@doi [\apj] {10.1086/168055}, \href
  {http://adsabs.harvard.edu/abs/1989ApJ...346..735B} {346, 735}

\bibitem[\protect\citeauthoryear{{Boschman}, {Cazaux}, {Spaans}, {Hoekstra}  \&
  {Schlath{\"o}lter}}{{Boschman} et~al.}{2015}]{2015A&A...579A..72B}
{Boschman} L.,  {Cazaux} S.,  {Spaans} M.,  {Hoekstra} R.,   {Schlath{\"o}lter}
  T.,  2015, \mn@doi [\aap] {10.1051/0004-6361/201323165}, \href
  {http://adsabs.harvard.edu/abs/2015A%26A...579A..72B} {579, A72}

\bibitem[\protect\citeauthoryear{{Candian} \& {Sarre}}{{Candian} \&
  {Sarre}}{2015}]{2015MNRAS.448.2960C}
{Candian} A.,  {Sarre} P.~J.,  2015, \mn@doi [\mnras] {10.1093/mnras/stv192},
  \href {http://adsabs.harvard.edu/abs/2015MNRAS.448.2960C} {448, 2960}

\bibitem[\protect\citeauthoryear{{Cernicharo}}{{Cernicharo}}{2004}]{2004ApJ...608L..41C}
{Cernicharo} J.,  2004, \mn@doi [\apjl] {10.1086/422170}, \href
  {http://adsabs.harvard.edu/abs/2004ApJ...608L..41C} {608, L41}

\bibitem[\protect\citeauthoryear{{Cerrigone}, {Hora}, {Umana}  \&
  {Trigilio}}{{Cerrigone} et~al.}{2009}]{2009ApJ...703..585C}
{Cerrigone} L.,  {Hora} J.~L.,  {Umana} G.,   {Trigilio} C.,  2009, \mn@doi
  [\apj] {10.1088/0004-637X/703/1/585}, \href
  {http://adsabs.harvard.edu/abs/2009ApJ...703..585C} {703, 585}

\bibitem[\protect\citeauthoryear{{Cherchneff}, {Barker}  \&
  {Tielens}}{{Cherchneff} et~al.}{1992}]{1992ApJ...401..269C}
{Cherchneff} I.,  {Barker} J.~R.,   {Tielens} A.~G.~G.~M.,  1992, \mn@doi
  [\apj] {10.1086/172059}, \href
  {http://adsabs.harvard.edu/abs/1992ApJ...401..269C} {401, 269}

\bibitem[\protect\citeauthoryear{{Cohen} \& {Barlow}}{{Cohen} \&
  {Barlow}}{2005}]{2005MNRAS.362.1199C}
{Cohen} M.,  {Barlow} M.~J.,  2005, \mn@doi [\mnras]
  {10.1111/j.1365-2966.2005.09366.x}, \href
  {http://adsabs.harvard.edu/abs/2005MNRAS.362.1199C} {362, 1199}

\bibitem[\protect\citeauthoryear{{Compi{\`e}gne}, {Abergel}, {Verstraete},
  {Reach}, {Habart}, {Smith}, {Boulanger}  \& {Joblin}}{{Compi{\`e}gne}
  et~al.}{2007}]{2007A&A...471..205C}
{Compi{\`e}gne} M.,  {Abergel} A.,  {Verstraete} L.,  {Reach} W.~T.,  {Habart}
  E.,  {Smith} J.~D.,  {Boulanger} F.,   {Joblin} C.,  2007, \mn@doi [\aap]
  {10.1051/0004-6361:20066172}, \href
  {http://adsabs.harvard.edu/abs/2007A%26A...471..205C} {471, 205}

\bibitem[\protect\citeauthoryear{{Cox} et~al.,}{{Cox}
  et~al.}{1998}]{1998ApJ...495L..23C}
{Cox} P.,  et~al., 1998, \mn@doi [\apjl] {10.1086/311212}, \href
  {http://adsabs.harvard.edu/abs/1998ApJ...495L..23C} {495, L23}

\bibitem[\protect\citeauthoryear{{Delgado-Inglada} \&
  {Rodr{\'{\i}}guez}}{{Delgado-Inglada} \&
  {Rodr{\'{\i}}guez}}{2014}]{2014ApJ...784..173D}
{Delgado-Inglada} G.,  {Rodr{\'{\i}}guez} M.,  2014, \mn@doi [\apj]
  {10.1088/0004-637X/784/2/173}, \href
  {http://adsabs.harvard.edu/abs/2014ApJ...784..173D} {784, 173}

\bibitem[\protect\citeauthoryear{{Frenklach} \& {Feigelson}}{{Frenklach} \&
  {Feigelson}}{1989}]{1989ApJ...341..372F}
{Frenklach} M.,  {Feigelson} E.~D.,  1989, \mn@doi [\apj] {10.1086/167501},
  \href {http://adsabs.harvard.edu/abs/1989ApJ...341..372F} {341, 372}

\bibitem[\protect\citeauthoryear{{Gail} \& {Sedlmayr}}{{Gail} \&
  {Sedlmayr}}{1988}]{1988A&A...206..153G}
{Gail} H.-P.,  {Sedlmayr} E.,  1988, \aap, \href
  {http://adsabs.harvard.edu/abs/1988A%26A...206..153G} {206, 153}

\bibitem[\protect\citeauthoryear{{Garc{\'{\i}}a-Hern{\'a}ndez} \&
  {G{\'o}rny}}{{Garc{\'{\i}}a-Hern{\'a}ndez} \&
  {G{\'o}rny}}{2014}]{2014A&A...567A..12G}
{Garc{\'{\i}}a-Hern{\'a}ndez} D.~A.,  {G{\'o}rny} S.~K.,  2014, \mn@doi [\aap]
  {10.1051/0004-6361/201423620}, \href
  {http://adsabs.harvard.edu/abs/2014A%26A...567A..12G} {567, A12}

\bibitem[\protect\citeauthoryear{{Garc{\'{\i}}a-Rojas}, {Pe{\~n}a}, {Morisset},
  {Delgado-Inglada}, {Mesa-Delgado}  \& {Ruiz}}{{Garc{\'{\i}}a-Rojas}
  et~al.}{2013}]{2013A&A...558A.122G}
{Garc{\'{\i}}a-Rojas} J.,  {Pe{\~n}a} M.,  {Morisset} C.,  {Delgado-Inglada}
  G.,  {Mesa-Delgado} A.,   {Ruiz} M.~T.,  2013, \mn@doi [\aap]
  {10.1051/0004-6361/201322354}, \href
  {http://adsabs.harvard.edu/abs/2013A%26A...558A.122G} {558, A122}

\bibitem[\protect\citeauthoryear{{Guzman-Ramirez}, {Zijlstra},
  {N{\'{\i}}chuim{\'{\i}}n}, {Gesicki}, {Lagadec}, {Millar}  \&
  {Woods}}{{Guzman-Ramirez} et~al.}{2011}]{2011MNRAS.414.1667G}
{Guzman-Ramirez} L.,  {Zijlstra} A.~A.,  {N{\'{\i}}chuim{\'{\i}}n} R.,
  {Gesicki} K.,  {Lagadec} E.,  {Millar} T.~J.,   {Woods} P.~M.,  2011, \mn@doi
  [\mnras] {10.1111/j.1365-2966.2011.18502.x}, \href
  {http://adsabs.harvard.edu/abs/2011MNRAS.414.1667G} {414, 1667}

\bibitem[\protect\citeauthoryear{{Guzman-Ramirez}, {Lagadec}, {Jones},
  {Zijlstra}  \& {Gesicki}}{{Guzman-Ramirez}
  et~al.}{2014}]{2014MNRAS.441..364G}
{Guzman-Ramirez} L.,  {Lagadec} E.,  {Jones} D.,  {Zijlstra} A.~A.,   {Gesicki}
  K.,  2014, \mn@doi [\mnras] {10.1093/mnras/stu454}, \href
  {http://adsabs.harvard.edu/abs/2014MNRAS.441..364G} {441, 364}

\bibitem[\protect\citeauthoryear{{Habart}, {Boulanger}, {Verstraete}, {Pineau
  des For{\^e}ts}, {Falgarone}  \& {Abergel}}{{Habart}
  et~al.}{2003}]{2003A&A...397..623H}
{Habart} E.,  {Boulanger} F.,  {Verstraete} L.,  {Pineau des For{\^e}ts} G.,
  {Falgarone} E.,   {Abergel} A.,  2003, \mn@doi [\aap]
  {10.1051/0004-6361:20021489}, \href
  {http://adsabs.harvard.edu/abs/2003A%26A...397..623H} {397, 623}

\bibitem[\protect\citeauthoryear{{Henney}, {Williams}, {Ferland}, {Shaw}  \&
  {O'Dell}}{{Henney} et~al.}{2007}]{2007ApJ...671L.137H}
{Henney} W.~J.,  {Williams} R.~J.~R.,  {Ferland} G.~J.,  {Shaw} G.,   {O'Dell}
  C.~R.,  2007, \mn@doi [\apjl] {10.1086/525023}, \href
  {http://adsabs.harvard.edu/abs/2007ApJ...671L.137H} {671, L137}

\bibitem[\protect\citeauthoryear{{Hony}, {Van Kerckhoven}, {Peeters},
  {Tielens}, {Hudgins}  \& {Allamandola}}{{Hony}
  et~al.}{2001}]{2001A&A...370.1030H}
{Hony} S.,  {Van Kerckhoven} C.,  {Peeters} E.,  {Tielens} A.~G.~G.~M.,
  {Hudgins} D.~M.,   {Allamandola} L.~J.,  2001, \mn@doi [\aap]
  {10.1051/0004-6361:20010242}, \href
  {http://adsabs.harvard.edu/abs/2001A%26A...370.1030H} {370, 1030}

\bibitem[\protect\citeauthoryear{{Hora}, {Marengo}, {Smith}, {Cerrigone}  \&
  {Latter}}{{Hora} et~al.}{2009}]{2009eimw.confE..29H}
{Hora} J.~L.,  {Marengo} M.,  {Smith} H.~A.,  {Cerrigone} L.,   {Latter} W.~B.,
   2009, in The Evolving ISM in the Milky Way and Nearby Galaxies. p.~29
  (\mn@eprint {arXiv} {0803.3937})

\bibitem[\protect\citeauthoryear{{Houck} et~al.,}{{Houck}
  et~al.}{2004}]{2004ApJS..154...18H}
{Houck} J.~R.,  et~al., 2004, \mn@doi [\apjs] {10.1086/423134}, \href
  {http://adsabs.harvard.edu/abs/2004ApJS..154...18H} {154, 18}

\bibitem[\protect\citeauthoryear{{Joblin}, {Szczerba}, {Bern{\'e}}  \&
  {Szyszka}}{{Joblin} et~al.}{2008}]{2008A&A...490..189J}
{Joblin} C.,  {Szczerba} R.,  {Bern{\'e}} O.,   {Szyszka} C.,  2008, \mn@doi
  [\aap] {10.1051/0004-6361:20079061}, \href
  {http://adsabs.harvard.edu/abs/2008A%26A...490..189J} {490, 189}

\bibitem[\protect\citeauthoryear{{Jones}, {Tielens}  \& {Hollenbach}}{{Jones}
  et~al.}{1996}]{1996ApJ...469..740J}
{Jones} A.~P.,  {Tielens} A.~G.~G.~M.,   {Hollenbach} D.~J.,  1996, \mn@doi
  [\apj] {10.1086/177823}, \href
  {http://adsabs.harvard.edu/abs/1996ApJ...469..740J} {469, 740}

\bibitem[\protect\citeauthoryear{{Kastner} et~al.,}{{Kastner}
  et~al.}{2012}]{2012AJ....144...58K}
{Kastner} J.~H.,  et~al., 2012, \mn@doi [\aj] {10.1088/0004-6256/144/2/58},
  \href {http://adsabs.harvard.edu/abs/2012AJ....144...58K} {144, 58}

\bibitem[\protect\citeauthoryear{{Langhoff}}{{Langhoff}}{1996}]{Langhoff1996}
{Langhoff} S.~R.,  1996, J. Phys. Chem, 100, 2819

\bibitem[\protect\citeauthoryear{{Liu}, {Liu}, {Barlow}  \& {Luo}}{{Liu}
  et~al.}{2004}]{2004MNRAS.353.1251L}
{Liu} Y.,  {Liu} X.-W.,  {Barlow} M.~J.,   {Luo} S.-G.,  2004, \mn@doi [\mnras]
  {10.1111/j.1365-2966.2004.08156.x}, \href
  {http://adsabs.harvard.edu/abs/2004MNRAS.353.1251L} {353, 1251}

\bibitem[\protect\citeauthoryear{{Manchado}, {Stanghellini}, {Villaver},
  {Garc{\'{\i}}a-Segura}, {Shaw}  \& {Garc{\'{\i}}a-Hern{\'a}ndez}}{{Manchado}
  et~al.}{2015}]{2015ApJ...808..115M}
{Manchado} A.,  {Stanghellini} L.,  {Villaver} E.,  {Garc{\'{\i}}a-Segura} G.,
  {Shaw} R.~A.,   {Garc{\'{\i}}a-Hern{\'a}ndez} D.~A.,  2015, \mn@doi [\apj]
  {10.1088/0004-637X/808/2/115}, \href
  {http://adsabs.harvard.edu/abs/2015ApJ...808..115M} {808, 115}

\bibitem[\protect\citeauthoryear{{Matsuura} et~al.,}{{Matsuura}
  et~al.}{2004}]{2004ApJ...604..791M}
{Matsuura} M.,  et~al., 2004, \mn@doi [\apj] {10.1086/382064}, \href
  {http://adsabs.harvard.edu/abs/2004ApJ...604..791M} {604, 791}

\bibitem[\protect\citeauthoryear{{Merino} et~al.,}{{Merino}
  et~al.}{2014}]{2014NatCo...5E3054M}
{Merino} P.,  et~al., 2014, \mn@doi [Nature Communications]
  {10.1038/ncomms4054}, \href
  {http://adsabs.harvard.edu/abs/2014NatCo...5E3054M} {5, 3054}

\bibitem[\protect\citeauthoryear{{Montillaud}, {Joblin}  \&
  {Toublanc}}{{Montillaud} et~al.}{2013}]{2013A&A...552A..15M}
{Montillaud} J.,  {Joblin} C.,   {Toublanc} D.,  2013, \mn@doi [\aap]
  {10.1051/0004-6361/201220757}, \href
  {http://adsabs.harvard.edu/abs/2013A%26A...552A..15M} {552, A15}

\bibitem[\protect\citeauthoryear{{Natta} \& {Hollenbach}}{{Natta} \&
  {Hollenbach}}{1998}]{1998A&A...337..517N}
{Natta} A.,  {Hollenbach} D.,  1998, \aap, \href
  {http://adsabs.harvard.edu/abs/1998A%26A...337..517N} {337, 517}

\bibitem[\protect\citeauthoryear{{O'Dell}, {Balick}, {Hajian}, {Henney}  \&
  {Burkert}}{{O'Dell} et~al.}{2002}]{2002AJ....123.3329O}
{O'Dell} C.~R.,  {Balick} B.,  {Hajian} A.~R.,  {Henney} W.~J.,   {Burkert} A.,
   2002, \mn@doi [\aj] {10.1086/340726}, \href
  {http://adsabs.harvard.edu/abs/2002AJ....123.3329O} {123, 3329}

\bibitem[\protect\citeauthoryear{{O'Dell}, {Sabbadin}  \& {Henney}}{{O'Dell}
  et~al.}{2007}]{2007AJ....134.1679O}
{O'Dell} C.~R.,  {Sabbadin} F.,   {Henney} W.~J.,  2007, \mn@doi [\aj]
  {10.1086/521823}, \href {http://adsabs.harvard.edu/abs/2007AJ....134.1679O}
  {134, 1679}

\bibitem[\protect\citeauthoryear{{O'Dell}, {Henney}  \& {Sabbadin}}{{O'Dell}
  et~al.}{2009}]{2009AJ....137.3815O}
{O'Dell} C.~R.,  {Henney} W.~J.,   {Sabbadin} F.,  2009, \mn@doi [\aj]
  {10.1088/0004-6256/137/4/3815}, \href
  {http://adsabs.harvard.edu/abs/2009AJ....137.3815O} {137, 3815}

\bibitem[\protect\citeauthoryear{{O'Dell}, {Ferland}, {Henney}  \&
  {Peimbert}}{{O'Dell} et~al.}{2013a}]{2013AJ....145...92O}
{O'Dell} C.~R.,  {Ferland} G.~J.,  {Henney} W.~J.,   {Peimbert} M.,  2013a,
  \mn@doi [\aj] {10.1088/0004-6256/145/4/92}, \href
  {http://adsabs.harvard.edu/abs/2013AJ....145...92O} {145, 92}

\bibitem[\protect\citeauthoryear{{O'Dell}, {Ferland}, {Henney}  \&
  {Peimbert}}{{O'Dell} et~al.}{2013b}]{2013AJ....145...93O}
{O'Dell} C.~R.,  {Ferland} G.~J.,  {Henney} W.~J.,   {Peimbert} M.,  2013b,
  \mn@doi [\aj] {10.1088/0004-6256/145/4/93}, \href
  {http://adsabs.harvard.edu/abs/2013AJ....145...93O} {145, 93}

\bibitem[\protect\citeauthoryear{{Pauzat}}{{Pauzat}}{2011}]{2011EAS....46...75P}
{Pauzat} F.,  2011, in {Joblin} C.,  {Tielens} A.~G.~G.~M.,  eds,  EAS
  Publications Series Vol. 46, EAS Publications Series. pp 75--93,
  \mn@doi{10.1051/eas/1146008}

\bibitem[\protect\citeauthoryear{{Pech}, {Joblin}  \& {Boissel}}{{Pech}
  et~al.}{2002}]{2002A&A...388..639P}
{Pech} C.,  {Joblin} C.,   {Boissel} P.,  2002, \mn@doi [\aap]
  {10.1051/0004-6361:20020416}, \href
  {http://adsabs.harvard.edu/abs/2002A%26A...388..639P} {388, 639}

\bibitem[\protect\citeauthoryear{{Perea-Calder{\'o}n},
  {Garc{\'{\i}}a-Hern{\'a}ndez}, {Garc{\'{\i}}a-Lario}, {Szczerba}  \&
  {Bobrowsky}}{{Perea-Calder{\'o}n} et~al.}{2009}]{2009A&A...495L...5P}
{Perea-Calder{\'o}n} J.~V.,  {Garc{\'{\i}}a-Hern{\'a}ndez} D.~A.,
  {Garc{\'{\i}}a-Lario} P.,  {Szczerba} R.,   {Bobrowsky} M.,  2009, \mn@doi
  [\aap] {10.1051/0004-6361:200811457}, \href
  {http://adsabs.harvard.edu/abs/2009A%26A...495L...5P} {495, L5}

\bibitem[\protect\citeauthoryear{{Pilleri}, {Montillaud}, {Bern{\'e}}  \&
  {Joblin}}{{Pilleri} et~al.}{2012}]{2012A&A...542A..69P}
{Pilleri} P.,  {Montillaud} J.,  {Bern{\'e}} O.,   {Joblin} C.,  2012, \mn@doi
  [\aap] {10.1051/0004-6361/201015915}, \href
  {http://adsabs.harvard.edu/abs/2012A%26A...542A..69P} {542, A69}

\bibitem[\protect\citeauthoryear{{Ricca}, {Bauschlicher}, {Boersma}, {Tielens}
  \& {Allamandola}}{{Ricca} et~al.}{2012}]{2012ApJ...754...75R}
{Ricca} A.,  {Bauschlicher} Jr. C.~W.,  {Boersma} C.,  {Tielens} A.~G.~G.~M.,
  {Allamandola} L.~J.,  2012, \mn@doi [\apj] {10.1088/0004-637X/754/1/75},
  \href {http://adsabs.harvard.edu/abs/2012ApJ...754...75R} {754, 75}

\bibitem[\protect\citeauthoryear{{Sahai}, {Morris}, {Werner}, {G{\"u}sten},
  {Wiesemeyer}  \& {Sandell}}{{Sahai} et~al.}{2012}]{2012A&A...542L..20S}
{Sahai} R.,  {Morris} M.~R.,  {Werner} M.~W.,  {G{\"u}sten} R.,  {Wiesemeyer}
  H.,   {Sandell} G.,  2012, \mn@doi [\aap] {10.1051/0004-6361/201219021},
  \href {http://adsabs.harvard.edu/abs/2012A%26A...542L..20S} {542, L20}

\bibitem[\protect\citeauthoryear{{Smith} et~al.,}{{Smith}
  et~al.}{2007}]{2007PASP..119.1133S}
{Smith} J.~D.~T.,  et~al., 2007, \mn@doi [\pasp] {10.1086/522634}, \href
  {http://adsabs.harvard.edu/abs/2007PASP..119.1133S} {119, 1133}

\bibitem[\protect\citeauthoryear{{Speck}, {Meixner}, {Jacoby}  \&
  {Knezek}}{{Speck} et~al.}{2003}]{2003PASP..115..170S}
{Speck} A.~K.,  {Meixner} M.,  {Jacoby} G.~H.,   {Knezek} P.~M.,  2003, \mn@doi
  [\pasp] {10.1086/345911}, \href
  {http://adsabs.harvard.edu/abs/2003PASP..115..170S} {115, 170}

\bibitem[\protect\citeauthoryear{{Tielens}}{{Tielens}}{2005}]{2005pcim.book.....T}
{Tielens} A.~G.~G.~M.,  2005, {The Physics and Chemistry of the Interstellar
  Medium}

\bibitem[\protect\citeauthoryear{{Vicini}, {Natta}, {Marconi}, {Testi},
  {Hollenbach}  \& {Draine}}{{Vicini} et~al.}{1999}]{1999A&A...342..823V}
{Vicini} B.,  {Natta} A.,  {Marconi} A.,  {Testi} L.,  {Hollenbach} D.,
  {Draine} B.~T.,  1999, \aap, \href
  {http://adsabs.harvard.edu/abs/1999A%26A...342..823V} {342, 823}

\bibitem[\protect\citeauthoryear{{Werner} et~al.,}{{Werner}
  et~al.}{2004}]{2004ApJS..154....1W}
{Werner} M.~W.,  et~al., 2004, \mn@doi [\apjs] {10.1086/422992}, \href
  {http://adsabs.harvard.edu/abs/2004ApJS..154....1W} {154, 1}

\bibitem[\protect\citeauthoryear{{Zhen}, {Castellanos}, {Paardekooper},
  {Ligterink}, {Linnartz}, {Nahon}, {Joblin}  \& {Tielens}}{{Zhen}
  et~al.}{2015}]{2015ApJ...804L...7Z}
{Zhen} J.,  {Castellanos} P.,  {Paardekooper} D.~M.,  {Ligterink} N.,
  {Linnartz} H.,  {Nahon} L.,  {Joblin} C.,   {Tielens} A.~G.~G.~M.,  2015,
  \mn@doi [\apjl] {10.1088/2041-8205/804/1/L7}, \href
  {http://adsabs.harvard.edu/abs/2015ApJ...804L...7Z} {804, L7}

\bibitem[\protect\citeauthoryear{{van Hoof} et~al.,}{{van Hoof}
  et~al.}{2010}]{2010A&A...518L.137V}
{van Hoof} P.~A.~M.,  et~al., 2010, \mn@doi [\aap]
  {10.1051/0004-6361/201014590}, \href
  {http://adsabs.harvard.edu/abs/2010A%26A...518L.137V} {518, L137}

\makeatother
\end{thebibliography}
}

\label{lastpage}
\end{document}